# A combinatorial approach for studying the effect of Mg concentration on precipitation in an Al-Cu-Li alloy

E. Gumbmann[1,2,3], F. De Geuser[1,2], A. Deschamps[1,2], W. Lefebvre[4], F. Robaut[5], C. Sigli[3]

[1] Université Grenoble Alpes, SIMAP, F-38000 Grenoble, France
[2] CNRS, SIMAP, F-38000 Grenoble, France
[3] Constellium Technology Center, CS 10027, 38341 Voreppe Cedex, France
[4] Groupe de Physique des Matériaux, UMR CNRS 6634, Normandie University, University of Rouen and INSA Rouen, 76801 St.Etienne du Rouvray, France
[5] Consortium des Moyens Technologiques Communs, Grenoble-INP, 38402 St. Martin d'Hères, France

**Abstract**
We apply a combinatorial approach to study the influence of Mg concentration on the precipitation kinetics in an Al-Cu-Li alloy using a diffusion couple made by linear friction welding. The precipitation kinetics is monitored in the composition gradient material using simultaneous space and time-resolved in-situ small-angle X-ray scattering measurements during ageing, and the strengthening of the precipitates is evaluated by micro-hardness profiles. This data provides an evaluation of the amount of Mg necessary to promote precipitation of the $T_1$-$Al_2CuLi$ phase.

Al-Cu-Li alloys are currently experiencing a strong interest due to their combination of low weight, high strength and high toughness suitable for aerospace applications. In the latest generation of alloys (such as commercialized as AIRWARE®), the main strengthening phase sought is the $T_1$ phase of bulk composition $Al_2CuLi$. The bulk structure of this equilibrium phase was resolved by Van Smaalen et al. [1]. In aluminium, it appears as extremely thin platelets on $\{111\}_{Al}$ planes with aspect ratio up to 50-100. The structure of this phase embedded in the Al matrix has been resolved in detail by Donnadieu et al. [2] and Dwyer et al. [3]. Since earlier studies, it is known that obtaining an efficient precipitation strengthening in this system requires the addition of dislocations [4,5] and of minor solute elements, out of which the most prominent ones are Mg and Ag [5–7]. These elements have been shown to be included in the composition of the nanoscale $T_1$ phase [8,9]. Recently, atom probe tomography has demonstrated that they segregate to the $T_1$/Al interface, and that the Mg atoms are involved in a co-clustering or co-precipitation with Cu, linked with dislocations, very early during the ageing process [10]. Now that the qualitative role of the minor solute elements on the precipitation of $T_1$ has been evidenced, there is a need to understand what solute content is necessary to obtain the desired effect. In particular, the effect of minor solute element concentration on the precipitation of Cu and Li may be strongly non-linear, and even non monotonous as there may be a competition for solute (especially Cu) from the added Mg atoms that may hinder the formation of $T_1$ instead of promoting it. A traditional alloy series fabrication with a discrete distribution of minor solute concentrations would be extremely cumbersome and may not evidence some of these non-linear events. Instead, we propose to use a combinatorial approach using compositional gradient materials [11], where the main alloying content is kept constant and one minor solute content is varied. This approach has already been used to study composition effects on precipitation hardening in Al-Cu-Mg alloys [12]. However in this study the composition gradient was created simply by inter-diffusion between two alloys, which restricted the dimension of the concentration gradient zone to a few 100 µm, and left at the interface a brittle oxide layer that prevented any mechanical loading of the gradient material. These characteristics are not suitable to the needs of our study for two reasons. First, because the study of $T_1$ precipitation requires the introduction of plastic deformation after the solution treatment, which would be impossible with the presence of an oxide layer within the inter-diffusion zone. Second, because we aim at characterizing the precipitation kinetics by small-angle X-ray scattering within the diffusion couple, using simultaneous time and space-resolved measurements, so that a full map of precipitation kinetics in the composition space is obtained in a single experiment. Such measurements require a relatively large extension of the composition gradient compared to the spatial resolution of the experiment, which cannot be obtained by interdiffusion alone. Therefore, we have chosen to join the alloys of different composition by linear friction welding, which has been shown to be an efficient method for materials of dissimilar composition [13], ensuring a planar interface between the two welded materials, free of any oxide layer (which has been expulsed by plastic flow during the welding process).

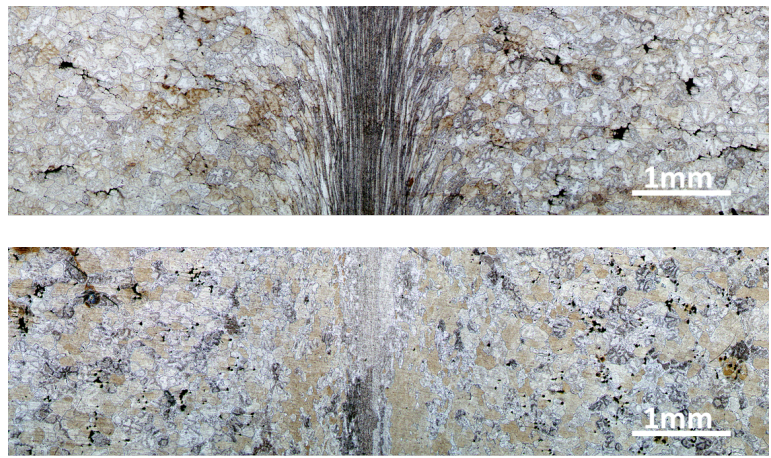

Figure 1: Optical micrographs of the linear friction weld (interface plane is vertical in the centre of the micrograph) in the as-welded state (top) and after the interdiffusion heat treatment (bottom).

The two studied alloys have the same base composition of 3.5 % Cu and 0.9 % Li (in wt). One of the two alloys has in addition 0.35 wt% Mg. The two alloys were joined as plates of thickness 25 mm by linear friction welding performed by Thompson Friction Welding. Figure 1 shows an optical micrograph of the interface between the two materials after welding and after the subsequent inter-diffusion treatment (see below). It is adequately planar and defect-free, so that the initial composition step between the two materials is well defined. Diffusion of Mg was activated by subjecting the joined couple to a homogenisation heat treatment at 515 °C for 14 days. Subsequent hot rolling was then used to further enlarge the width of the diffusion gradient. The pre-heating temperature was 500 °C and the thickness reduction from 6.5 to 1.8 mm was performed by a single rolling step. The evolution of the composition gradient, measured by electron probe micro analysis (EPMA) (20 keV, 300 nA) during these different steps is shown in Figure 2. In all cases the Cu content remains constant in the couple, whereas the Mg gradient, initially a sharp step, ends with a 10 mm length. One can emphasize at this stage that such a large characteristic length of the diffusion gradient ensures that, at the scale of the precipitation process (100 nm at most) the local composition of the alloy can be considered as constant so that no coupling between the macroscopic solute diffusion and the microscopic precipitation mechanisms needs to be considered.

Following the hot rolling step, a solution treatment was performed at 505°C for 30 min, using a slow heating ramp in order to avoid recrystallization. Following a quench in cold water, a pre-deformation of 4 % plastic strain was introduced. The tensile straining was performed normal to the original weld plane interface (materials loaded in series). It has been checked previously that the addition of 0.35% Mg to this alloy did not change significantly the flow stress of the Al-Cu-Li alloy in the as-quenched state so that the plastic strain (measured by an extensometer across the diffusion zone) was homogeneously distributed for all compositions. The material was kept for 3 days of natural aging before artificial aging was performed. The precipitation aging treatment included a ramp heating of 20 °C/h to 155 °C and further isothermal aging at 155 °C for several days.

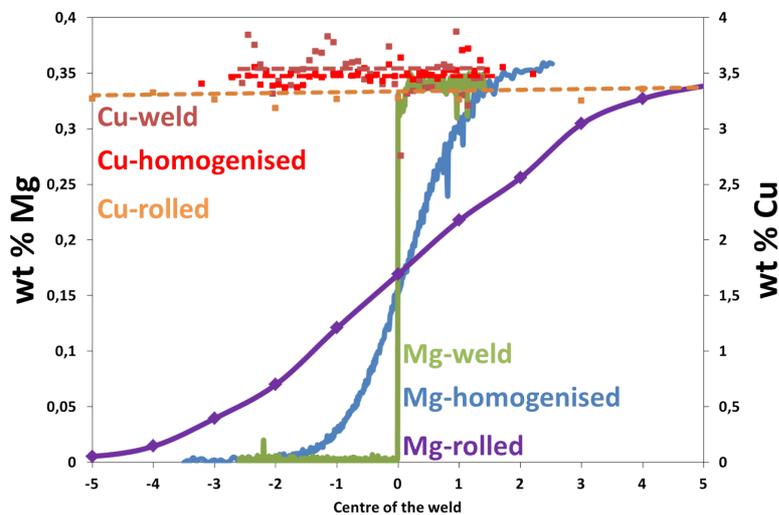

Figure 2: Diffusion gradient measured by EPMA after linear friction welding (green), after 14 days homogenisation treatment (blue) and after rolling (purple).

Precipitation kinetics was measured by small-angle X-ray scattering (SAXS). This method has been shown to be able to quantify in-situ the kinetics of $T_1$ phase precipitation in several recent studies [14–16]. It has also been shown to be capable of mapping heterogeneous microstructures in friction stir welds [17]. In the present case, we have combined spatial and time resolutions to perform a simultaneous evaluation of precipitation kinetics in the whole diffusion couple. SAXS measurements were performed on a laboratory rotating anode at the Cu Kα wavelength of 0.154 nm (8.048 keV), with a beam size diameter at the sample of 1mm. The SAXS set-up allowed measurements in a range of scattering vectors of 0.06 – 0.4 Å$^{-1}$. The aging treatment was conducted for more than 100h of ageing in a specially designed furnace, which was placed inside the sample vacuum chamber. By displacing the sample in between each measuring point, the complete composition gradient was scanned. A 2D DECTRIS PILATUS® photon counter camera was used for data acquisition and the counting time was 100 s. 18 measurements points, separated by 1 mm steps, were acquired across the sample, allowing a time resolution for a given position in the couple (corresponding to a given Mg concentration) of 30 min. The acquired 2D scattering data was converted to intensity vs. scattering vector by radial averaging [18]. In parallel, micro-hardness measurements were performed across a diffusion couple aged for different times in an oil bath. Measurements were made with a distance of 1 mm between each measurement with a Buehler Tukon™1102 automatic hardness machine using an applied load of 500 g with an indentation time of 10 seconds.

**Hardness measurements**

Figure 3a shows the evolution of hardness across the diffusion gradient measured after different ageing times at 155°C (EoR corresponds to the end of the heating ramp to 155°C). There is a clear effect of Mg concentration on the hardness distribution in the composition gradient, at all stages of the ageing treatment. At the end of the ageing treatment (the maximum hardness stays constant for several hours of artificial aging and resembles a plateau), a difference of 20 HV is observed between the Mg and non-Mg containing alloys. This difference is enhanced to almost 40 HV at intermediate times (12h at 155°C), showing that Mg promotes an acceleration of precipitation kinetics in the Al-Cu-Li alloy. The effect of Mg concentration is better evidenced by plotting the evolution of hardness vs. time for a series of Mg concentrations present in the diffusion couple (Figure 3b). This figure shows that the hardening kinetics is indeed more sluggish in the Mg free Al-Cu-Li alloy as compared to the Mg-containing alloys. Looking at the alloys with highest and lowest Mg-content, namely AlCuLi0.35Mg and AlCuLi, in Figure 3b, it reveals that the initial hardening is faster for AlCuLiMg than for AlCuLi. Furthermore, maximum hardness for AlCuLiMg is reached after 15 hours, whereas it takes up to 50 hours at 155 °C for AlCuLi to reach maximum hardness. Moreover, the effect of Mg content on the precipitation kinetics is not linear. Above 0.1%Mg, the hardening kinetics is less dependent on Mg concentration therefore demonstrating saturation behaviour.

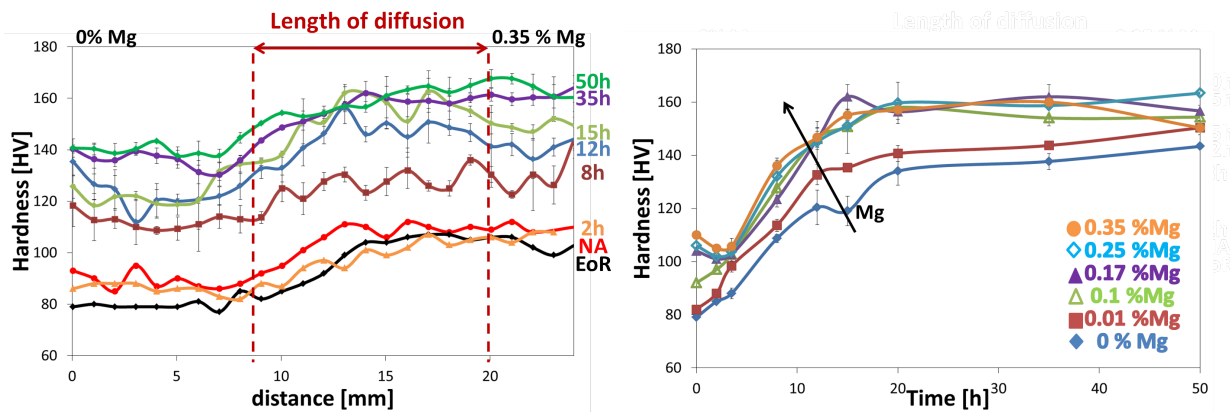

Figure 3: (a) Micro-hardness profiles across the diffusion gradient for different ageing times at 155°C; (b) evolution of micro-hardness with ageing time for a series of Mg concentrations characterized by their position within the diffusion gradient.

**SAXS measurements**

An evaluation of the volume fraction of precipitates was obtained from the SAXS data in the following way. First, it was assumed that no precipitates were present at the end of the heating ramp (which corresponds also to minimum hardness). Second, the volume fraction was assumed to be stabilized after 100h of ageing based on earlier studies [15,16]. In between, the relative volume fraction was calculated, proportional to the integrated intensity $Q_0$ [18]. One should remain careful about the interpretation of this relative volume fraction, because changes of composition may promote changes of precipitate types in this complex system; in particular, the absence of Mg may promote formation of some θ' phase at the expense of the $T_1$ phase, with a resulting different X-ray contrast. With this in mind, Figure 4a shows the evolution of relative volume fraction with ageing time for different Mg concentrations in the diffusion gradient. Since the diffusion gradient covers a length of 10 mm, but 18 mm were measured, several positions in the beginning and at the end of the gradient have the same composition. Similarly to the micro-hardness results, a strong effect of Mg concentration is evidenced, with faster precipitation kinetics in the Mg-rich regions. Around a Mg concentration of 0.1-0.2 %, the transition between slow and fast precipitation kinetics is observed, which is also consistent with the change observed in the hardening kinetics. Figure 4b shows together the hardness and volume fraction measurements for the first 50 hours of ageing. The two results do correspond particularly well. In a recent study, Dorin et al. [16] have actually shown that the relationship between $T_1$ precipitate volume fraction and yield strength is almost linear, although the detail of the relationship is more complicated.

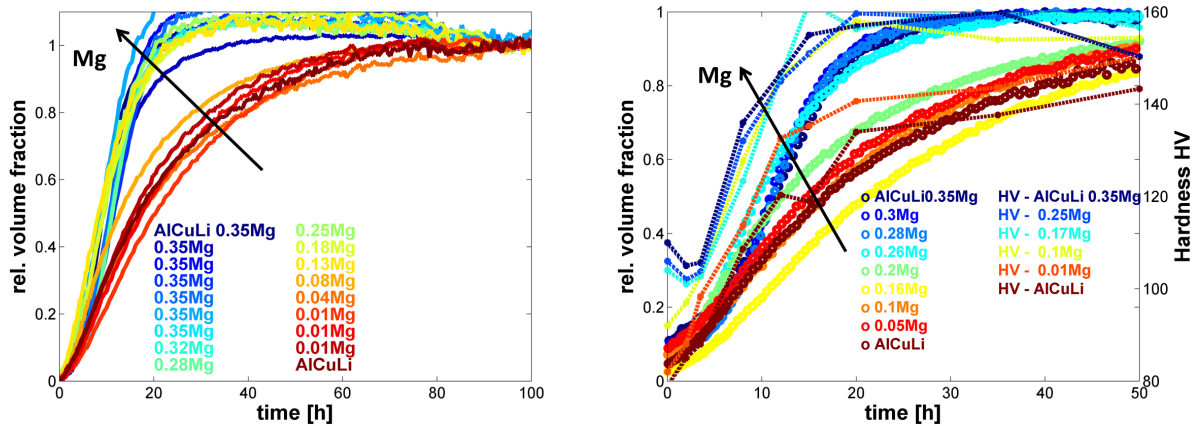

Figure 4: (a) Evolution of relative precipitate volume fraction measured from the SAXS data as a function of ageing time at 155°C for a series of Mg concentrations characterized by their position within the diffusion gradient; (b) correspondence between this precipitation kinetics during the first 50 hours and micro-hardness (symbols and dashed lines).

In conclusion, we have demonstrated a methodology for mapping the precipitation kinetics continuously in a material containing a composition gradient. This methodology involves the fabrication of a diffusion couple of sufficient dimensions using Linear Friction Welding followed by inter-diffusion heat treatment and hot rolling. This material is then subjected to an ageing treatment in a SAXS apparatus, where simultaneous space and time resolved measurements allow the precipitation kinetics characterization for a large number of compositions across the gradient. This methodology has been applied to evaluate the effect of Mg concentration on the precipitation kinetics and strengthening of an Al-Cu-Li alloy. We evidence that the addition of Mg accelerates the kinetics of precipitation and increases the strengthening capability. However the influence of Mg concentration on the precipitation kinetics is more pronounced for low additions. Our results help to determine the minimum amount of Mg necessary to observe this effect. More detailed studies are under way to better understand the effect of Mg on the precipitation mechanisms and will be published elsewhere.